\documentstyle[12pt,a4wide]{article}
\newcommand\as{\alpha_S}
\newcommand\gs{g_S}
\newcommand\MeV{{\,\sl MeV}}
\newcommand\GeV{{\,\sl GeV}}
\newcommand\slv{v\kern-5pt\raise1pt\hbox{$\scriptstyle/$}\kern1pt}
\newcommand\spur{\mathop{\rm Tr}\nolimits}

\begin{document}

\thispagestyle{empty}
\begin{flushright}
MZ-TH/97-19\\
WUE-ITP-97-014\\
hep-ph/9705447\\
May 1997\\
\end{flushright}
\vspace{0.5cm}
\begin{center}
{\Large\bf An Analysis of Diagonal and Non-diagonal}\\[.3cm]
{\Large\bf QCD Sum Rules for Heavy Baryons}\\[.3cm]
{\Large\bf at Next-to-Leading Order in \boldmath{$\as$}}\\
\vspace{1cm}
{\large S. Groote,\footnote{Groote@dipmza.physik.uni-mainz.de}
J.G. K\"orner\footnote{Koerner@dipmza.physik.uni-mainz.de}}\\[.5cm]
Institut f\"ur Physik, Johannes-Gutenberg-Universit\"at,\\[.2cm]
Staudinger Weg 7, D-55099 Mainz, Germany\\[1cm]
{\large O.I. Yakovlev\footnote{%
O\hbox to5pt{\hrulefill}Yakovlev@physik.uni-wuerzburg.de;\\
on leave from Budker Institute of Nuclear Physics (BINP),\\
pr. Lavrenteva 11, Novosibirsk, 630090, Russia}}\\[.5cm]
Institut f\"ur theoretische Physik II,\\[.2cm]
Bayrische Julius-Maximilians-Universit\"at,\\[.2cm]
Am Hubland, D-97074 W\"urzburg, Germany
\vspace{1cm}
\end{center}

\begin{abstract}
We consider diagonal and non-diagonal QCD sum rules for the ground state 
heavy baryons to leading order in $1/m_Q$ and at next-to-leading order in 
$\as$. In the non-diagonal case we evaluate the eight different two-loop 
diagrams which determine the perturbative $\as$-corrections to the Wilson 
coefficient of the quark condensate in the Operator Product Expansion. The 
QCD corrections to the non-diagonal sum rules are moderate compared to the 
QCD corrections in the diagonal case. We also consider constituent type 
sum rules using constituent type interpolating currents. The obtained 
results are in reasonable agreement with the corresponding results obtained 
in the diagonal case. As central values for the bound state energies we 
find $m(\Lambda_Q)-m_Q\simeq 760\MeV$ and $m(\Sigma_Q)-m_Q\simeq 940\MeV$. 
The central values for the residues are given by 
$F(\Lambda_Q)\simeq 0.030\GeV^3$ and $F(\Sigma_Q)\simeq 0.038\GeV^3$.

\end{abstract}

\newpage

\section{Introduction}
The knowledge of the non-perturbative properties of heavy hadrons such as 
their binding energies or their weak transition matrix elements are of 
fundamental importance for the determination of the fundamental parameters 
of the Standard Model. Among these are the quark masses and the values
of the Cabibbo-Kobayashi-Maskawa matrix elements. A convenient access to 
the properties of heavy hadrons containing one heavy quark is given by   
the Heavy Quark Effective Theory (HQET) which provides a systematic power
series expansion of physical matrix elements involving heavy hadrons in
terms of the inverse of the heavy quark mass (see for example~\cite{Neub}). 
While the case of heavy meson systems has been analyzed in great detail, 
corresponding calculations for heavy baryon systems have been lagging 
behind. This is unfortunate since results from the analysis of heavy baryon 
systems are expected to provide important supplementary information on the 
non-perturbative dynamics of QCD and on the fundamental parameters of the 
Standard Model. The importance of further theoretical studies on heavy 
baryon systems is highlighted by the fact that there is now an abundance of 
new experimental data on heavy baryon decays sparked by recent advances in 
microvertexing techniques. This data needs to be analyzed and interpreted 
theoretically.

A convenient and well-trusted tool to investigate the non-perturbative
properties of heavy hadrons is the QCD sum rule method~\cite{Vain}. The 
first application of the QCD sum rule method to heavy baryons was considered 
some time ago by Shuryak~\cite{Shur} who studied heavy baryons in the static 
limit given by the leading term in the $1/m_Q$ expansion. The work of 
Shuryak~\cite{Shur} was revised and extended in~\cite{GrYa,BaCha}. An 
analysis of heavy baryons containing large but finite quark masses $m_Q$ was 
undertaken in~\cite{Blok,BaDo,Cher}.

We have recently been embarking on a program to improve on previous analysis 
of heavy baryon sum rules by including first order radiative corrections in 
the analysis. In~\cite{GKY} we determined the two-loop anomalous dimensions 
of the static heavy baryon currents. In~\cite{GKY1} we determined the 
perturbative $\as$ corrections to the leading, dimension zero, term in the 
operator product expansion (OPE) of the static heavy baryon current 
correlator. Similar to the heavy meson case investigated e.g.\ 
in~\cite{BrGr1,BaBa} the radiative corrections to the perturbative 
dimension zero term are quite large. The results of~\cite{GKY,GKY1} were 
used to construct and analyze so-called diagonal QCD sum rules for heavy 
baryons~\cite{GKY1}.

Here the term ``diagonal'' refers to a particular feature of heavy baryon
currents and their current products. For every baryonic state there are two 
independent interpolating currents even in the static 
limit~\cite{Shur,GrYa,GKY}. One can thus construct diagonal sum rules from 
current correlators of the same baryon current and non-diagonal sum rules 
from current correlators of different baryon currents. The structure of the 
sum rules for the two cases is qualitatively quite different. Nevertheless, 
they must be considered on the same footing. In~\cite{GKY1} we provided a 
detailed analysis of the diagonal sum rules. The main part of the present 
paper is devoted to an analysis of the non-diagonal sum rules including 
radiative corrections. We compare our results with those obtained from the 
analysis of the diagonal sum rules. Using results from~\cite{GKY1} we also 
analyze mixed sum rules where we use constituent type current combinations 
in the current correlators.  

In order to provide a brief synapsis of the structural differences 
of the diagonal and non-diagonal sum rules let us briefly recapitulate
the main features of the diagonal sum rule analysis~\cite{Shur,GrYa,GKY1}:

\begin{itemize}

\item QCD sum rules based on diagonal correlators feature a leading order 
spectral density which grows rapidly with energy as 
$\rho(\omega)\sim\omega^5$. This rapid growth introduces a strong dependence 
of the results on the assumed value of the continuum threshold.

\item The QCD radiative correction to the leading order spectral density 
amounts to about $100\%$ at the renormalization scale $\mu=1\GeV$.

\item  Despite of the large QCD corrections to the Wilson coefficients in
the OPE, the lowest order sum rules and the radiatively corrected sum rules 
predict nearly the same values for the masses and the residues, while the 
stability region of the sum rule results appears at slightly shifted values
of the continuum threshold.  

\end{itemize}

It is clear that one also needs to analyze the non-diagonal sum rules
in addition to the diagonal sum rules if only for reasons of consistency. 
A welcome property of the non-diagonal correlator is the quite ``normal'' 
behaviour $\rho(\omega)\sim\omega^2$ of the spectral density and the fact 
that the QCD corrections are moderate. 

The paper is organized as follows. In Sect.~2 we introduce our notation 
and construct the correlator of two heavy baryon currents. We also recall 
the form of the known QCD corrections to the dimension zero term of the OPE. 
In Sect.~3 we present our results on the QCD corrections to the dimension 
three contribution in the OPE, which is proportional to the vacuum expectation 
value of the product of the quark and the antiquark field. We also construct 
generalized QCD sum rules which incorporate both the diagonal and the 
non-diagonal case. Sect.~4 contains the results of our numerical analysis, 
our final numbers and a discussion of the results. In an Appendix we collect 
our results on the evaluation of the radiative two-loop corrections to the 
dimension-three condensate contribution. The results are given for 
$D$-dimensional space-time using the most general baryon current structure.

\section{Correlator of two baryonic currents} 
\subsection{Basic notions}
In this section we briefly recapitulate the basic notions involved in the 
construction of QCD sum rules for heavy baryons. This also serves to 
introduce our notation which closely follows the one used in~\cite{GKY,GKY1}. 
The starting point is given by the correlator of two baryonic currents
($i,j=1,2$)
\begin{equation}\label{correlator}
\Pi_{ij}(\omega=k\cdot v)
  =i\int d^4xe^{ikx}\langle 0|T\{J_i(x),\bar J_j(0)\}|0\rangle,
\end{equation}
where $k_\mu$ is the residual momentum of the heavy quark and $v_\mu$ is 
the four-velocity of the heavy baryon, the product of both being denoted by 
$\omega$. The residual momentum and the four-velocity are related by 
$p_\mu=m_Qv_\mu+k_\mu$, where $p_\mu$ denotes the momentum of the heavy 
quark and $m_Q$ is its mass. As was mentioned before, there are two possible 
choices of interpolating currents for each of the heavy baryon states. 
Neglecting the flavour and colour structure for the moment, these are given 
by\footnote{Here we use a rather symbolic notation. The Dirac strings 
$\Gamma$ and $\Gamma'$ can carry Lorentz indices. A contraction on the 
Lorentz indices is always implied when writing the currents in the form of 
Eq.~(\ref{currents})}
\begin{equation}\label{currents}
J_1=[q^{T}C\Gamma_1 q]\Gamma'Q\qquad\mbox{and}\qquad
J_2=[q^{T}C\Gamma_2q]\Gamma^{'}Q,
\end{equation}
where
\begin{eqnarray}\label{twoDirac}
\Gamma_1=\Gamma\quad\mbox{and}\quad\Gamma_2=\Gamma\slv.
\end{eqnarray}
To be more precise, it is clear that the multiplication of the light-side 
Dirac structure $\Gamma$ with $\slv$ does not change the quantum numbers of 
the interpolating current but does change the structure of the current. 
Multiplying the heavy-side structure $\Gamma'$ with $\slv$, however, does 
not change the structure of the interpolating current since $\slv Q=Q$ in 
the static limit.

In the static limit one has two types of heavy ground state baryons 
depending on whether the light diquark system is in a spin~$0$ or in a 
spin~$1$ state. We shall employ a generic notation and refer to the first 
type (spin~$0$ diquark) as $\Lambda_Q$-type heavy baryons. The Dirac 
structure of the interpolating current is given by $\Gamma=\gamma_5$ and 
$\Gamma'=1$ in this case. In the second case (spin~$1$ diquark) one has a 
doublet of degenerate $\Sigma_Q$-type states with overall spin~$1/2$ 
and~$3/2$. For the spin~$1/2$ $\Sigma_Q$-type state the interpolating 
current is given by $\Gamma=\gamma^\mu_\perp\equiv\gamma^\mu-\slv v^\mu$ and 
$\Gamma'=\gamma^\mu_\perp\gamma_5$. The explicit form of the spin~$3/2$ 
interpolating current ($\Sigma_Q^*$-type state) can be found e.g.\ 
in~\cite{GKY1}.

For a general analysis it proves convenient to represent the general 
light-side Dirac structure of the currents in Eq.~(\ref{currents}) by an 
antisymmetrized product of $n$ Dirac matrices 
$\Gamma=\gamma^{[\mu_1}\cdots\gamma^{\mu_n]}$. 
When calculating the one- and two-loop vertex corrections to the baryon 
currents in Eq.~(\ref{currents}) one encounters $\gamma$-contractions of 
the form $\gamma_\alpha\Gamma\gamma^\alpha$ and $\slv\Gamma\slv$. The 
$\gamma_\alpha$-contraction leads to an $n$-dependence according to
\begin{equation}
\label{Dirac1}
\gamma_\alpha\Gamma\gamma^\alpha=(-1)^n(D-2n)\Gamma,
\end{equation}
where $D$ denotes the space-time dimension. The $\slv$-contraction depends 
in addition on the parameter $s$ which takes the value ($s=+1$) and ($s=-1$) 
for an even or odd number of $\slv$'s in $\Gamma$, respectively. The 
$\slv$-contraction reads
\begin{equation}\label{Dirac2}
\slv\Gamma\slv=(-1)^ns\Gamma.
\end{equation}
Some of our results in the next sections are given in terms of the most 
general Dirac structure of heavy baryon currents involving the parameters 
$n$ and $s$ whose definitions should be kept in mind. For the convenience 
of the reader we list the relevant $(n,s)$-values for the cases studied in 
this paper in Table~\ref{tab1}.
\begin{table}\begin{center}
\begin{tabular}{|l|c|c|l|}\hline
$\Gamma$&$n$&$s$&particles\\\hline\hline
$\gamma_5$&$0$&$+1$&$\Lambda_1$\\\hline
$\gamma_5\gamma_0$&$1$&$-1$&$\Lambda_2$\\\hline
$\vec\gamma$&$1$&$+1$&$\Sigma_1,\Sigma^*_1$\\\hline
$\gamma_0\vec\gamma$&$2$&$-1$&$\Sigma_2,\Sigma^*_2$\\\hline
\end{tabular}
\caption{\label{tab1}Specific values of the parameter pair $(n,s)$ for 
particular cases of the light-side Dirac structure $\Gamma$.}
\end{center}\end{table}

\subsection{Anomalous dimensions of heavy baryon currents}
The one-loop and two-loop renormalization of the static heavy baryon
currents and their anomalous dimensions were considered in~\cite{GrYa} 
and~\cite{GKY}, respectively. Note that the anomalous dimensions of the 
currents differ in general from those in conventional QCD. We define the 
anomalous dimensions in terms of the expansion 
$\gamma=\sum_k(\as/4\pi)^k\gamma_k$. The one-loop anomalous dimension for 
the general current case depends only on the parameter $n$ which specifies 
the light-side Dirac structure. The one-loop anomalous dimension is given 
by~\cite{GrYa,GKY}
\begin{equation}\label{oneloopAD}
\gamma_1=-\frac43((n-2)^2+2).
\end{equation}
The general $(n,s)$-dependent formula for the two-loop anomalous dimension 
case is rather lengthy and will therefore not be listed here. The general
formula can be found in~\cite{GKY}. Here we specify to the case of the 
heavy ground state baryons and give the expansion to two-loop order using the 
$\overline{\rm MS}$-scheme and a naively anticommuting $\gamma_5$. One has
\begin{eqnarray}
\gamma_{\Lambda1}&=&-8\left(\frac{\as}{4\pi}\right)
  +\underbrace{\frac19(16\zeta(2)+40N_f-796)}_{\approx-72.19}
  \left(\frac{\as}{4\pi}\right)^2,\label{andimlam1}\\ 
\gamma_{\Lambda2}&=&-4\left(\frac{\as}{4\pi}\right)
  +\underbrace{\frac19(16\zeta(2)+20N_f-322)}_{\approx-26.19}
  \left(\frac{\as}{4\pi}\right)^2,\label{andimlam2}\\
\gamma_{\Sigma1}&=&-4\left(\frac{\as}{4\pi}\right)
  +\underbrace{\frac19(16\zeta(2)+20N_f-290)}_{\approx-22.63}
  \left(\frac{\as}{4\pi}\right)^2,\\
\gamma_{\Sigma2}&=&-\frac83\left(\frac{\as}{4\pi}\right)
  +\underbrace{\frac1{27}(48\zeta(2)+8N_f+324)}_{\approx 15.81}
  \left(\frac{\as}{4\pi}\right)^2,
\end{eqnarray}
where the numerical values are given for the case of three light flavours 
($N_f=3$).

\section{Diagonal, non-diagonal and mixed correlators}
As mentioned before, the two independent currents give rise to two 
independent types of correlators, namely the diagonal correlators 
$\langle J_1\bar J_1\rangle$ and $\langle J_2\bar J_2\rangle$, and the 
non-diagonal correlators $\langle J_1\bar J_2\rangle$ and 
$\langle J_2\bar J_1\rangle$. In the general case, one may even consider 
correlators built from a linear combination $J=aJ_1+(1-a)J_2$ of currents 
with an arbitrary coefficient $a$ ($0\le a\le 1$). We shall, however, not 
discuss the most general linear combination of currents in this paper. 
Later on we investigate the case $a=1/2$. The choice $J=(J_1+J_2)/2$ 
corresponds to a constituent quark model current which has maximal overlap 
with the ground state baryons in the constituent quark model picture.

Following the standard QCD sum rule method~\cite{Vain}, the correlator is 
calculated in the Euclidean region $-\omega\approx 1-2\GeV$ including 
perturbative and non-perturbative contributions. In the Euclidean region 
the non-perturbative contributions are expected to form a convergent series. 
The non-perturbative effects are taken into account by employing an OPE for 
the time-ordered product of the currents in Eq.~(\ref{correlator}). One 
then has 
\begin{eqnarray}
\langle T\{J(x),\bar J(0)\}\rangle &=&\sum_d C_d(x^2)O_d \nonumber \\
&=&C_0(x^2)O_0+C_3(x^2)O_3+C_4(x^2)O_4+C_5(x^2)O_5+\ldots
\end{eqnarray}
where the $O_d$ are vacuum expectation values of local operators whose 
dimensions are labelled by their subscripts $d$. $O_0=\hat 1$ corresponds 
to the so called perturbative term, $O_3=\langle\bar qq\rangle$ is a quark 
condensate term, $O_4=\as\langle G^2\rangle$ is a gluon condensate term, 
$O_5=\gs\langle\bar q\sigma_{\mu\nu}G^{\mu\nu}q\rangle$ is a mixed 
quark-gluon condensate, and so on. The expansion coefficients $C_d(x^2)$ 
are the associated coefficient functions or Wilson coefficients of the OPE.

A straightforward dimensional analysis shows that the OPE of the diagonal
correlators $\langle J_1\bar J_1\rangle$ and $\langle J_2\bar J_2\rangle$ 
contains only even-dimensional terms, while the OPE of the non-diagonal 
correlators contains only odd-dimensional terms. This classification is 
preserved when radiative corrections are included, assuming the light 
quarks to be massless. We apply radiative corrections only to the leading 
terms in the OPE because the non-leading contributions are small.

The diagonal case was studied in detail in~\cite{GKY1}. It was shown that 
the QCD corrections to the spectral density of the correlator function 
$P(\omega)$ are quite large. It is quite intriguing that the contributions 
of the four different three-loop diagrams that contribute to the 
perturbative dimension zero piece shown in Fig.~1 can be collected into one 
compact formula~\cite{GKY1}:    
\begin{eqnarray}
\frac{\rho_0^{\rm QCD}(\omega,\mu)}{\rho_0^{\rm Born}(\omega)}
  =1+\frac{\as}{4\pi}\Bigg[\ln\left(\frac\mu{2\omega}\right)
  \frac83(n^2-4n+6)+\frac8{45}(60\zeta(2)+38n^2-137n+273)\Bigg].
\end{eqnarray}
The number $n$ specifies the light-side Dirac structure of the baryon 
currents as before. Note that the coefficient of the logarithmic term 
coincides with the one-loop anomalous dimension of the diagonal correlator 
which in this case is equal to two times the anomalous dimension of the 
baryon current itself.

\subsection{Non-diagonal correlators}
The non-diagonal correlator of the two heavy baryon currents reads
\begin{equation}\label{OPE}
\Pi_{12}(\omega)
  =i\int d^4x\,e^{ikx}\langle 0|T\{J_1(x)\bar J_2(0)\}|0\rangle
  =\Gamma_1'\frac{1+\slv}2\bar\Gamma_2'\frac14\spur(\Gamma_1\bar\Gamma_2)
  P_{12}(\omega).
\end{equation}
We have suppressed the flavour and colour labels in Eq.~(\ref{OPE}). The 
OPE for the non-diagonal correlator contains a term 
$O_3(\mu)=\langle\bar qq\rangle$ proportional to the quark condensate, a 
mixed quark-gluon condensate term 
$O_5(\mu)=\gs\langle\bar q\sigma_{\mu\nu}G^{\mu\nu}q\rangle
\equiv m^2_0\langle\bar qq\rangle$, a term
$O_7(\mu)=\langle\bar qq\rangle\langle\as G^2\rangle$, and a term 
$O_9=\as\langle\bar qq\rangle^3$. Taking into account 
these four condensate contributions, the Fourier transform of the scalar 
correlator function $P(\omega)$ is given by
\begin{eqnarray}\label{OPEnondiag}
P(t)&=&P_{\rm OPE}(t)\\
  &=&-i\frac{2\theta(t)}{\pi^2t^3}\left(O_3(\mu)
  +\frac{t^2}{16}\Big(1-\frac c2\Big)O_5(\mu)
  +\frac{\pi t^4}{288}\Big(1-\frac c2\Big)O_7(\mu)
  -\frac{\pi^3t^6}{972}O_9(\mu)\right),\nonumber
\end{eqnarray}
where $c$ is a Clebsch-Gordan type factor which takes the values $c=1$ for
the $\Lambda_Q$-type and $c=-1/3$ for the $\Sigma_Q$-type doublet 
$\{\Sigma_Q,\Sigma_Q^*\}$ ground state baryons. The correlator function 
$P_{\rm OPE}(\omega)$ satisfies the 
dispersion relation
\begin{equation}\label{dispersion}
P_{\rm OPE}(\omega)=P(\omega)
  =\int_0^\infty\frac{\rho(\omega')d\omega'}{\omega'-\omega-i0}
  +\hbox{\rm subtraction},
\end{equation}
where $\rho(\omega)={\sl Im}(P(\omega))/\pi$ is the spectral density of 
the scalar correlation function. Taking into account the above four 
condensate contributions, the lowest order spectral density of the first 
two contributions is given by~\cite{GrYa,BaCha} 
\begin{equation}\label{rho3rho5}
\rho_3(\omega)=-\frac{\langle\bar qq\rangle}{\pi^2}\omega^2
  \quad\mbox{and}\quad
\rho_5(\omega)=2\left(1-\frac c2\right)
  \frac{\langle\bar qq\rangle m^2_0}{16\pi^2}.
\end{equation}

Next we compute the radiative corrections to the quark condensate term 
$\rho_3(\omega)$. There are altogether eight different contributing 
diagrams which are shown in Fig.~2. Their contributions were evaluated with 
the help of the algorithm developed in~\cite{BrGr}. As a check on the 
calculation we used a general covariant gauge for the gluon. The gauge 
dependence was found to drop out in the sum of the contributions. 

Collecting together the one- and two-loop contributions to the dimension 
three scalar correlation function one has
\begin{equation}\label{perturb}
P_3(\omega)=-\frac{\langle\bar qq\rangle}{2\pi^2}\omega^2
  \left(\left(\frac{-2\omega}\mu\right)^{D-4}C_0D_0
  +\frac{\gs^2}{(4\pi)^{D/2}}
  \left(\frac{-2\omega}\mu\right)^{2D-8}\sum_{i=1}^8C_iD_i\right),
\end{equation}
where $D=4-2\epsilon$ is the space-time dimension. There are a number of
colour factors in Eq.~(\ref{perturb}) the values of which are given by 
$C_0=N_c!$, $C_1=C_2=C_3=C_4=C_5=-N_c!C_B$ and $C_6=C_7=C_8=N_c!C_F$, where 
$N_c$ is the number of colours, $C_F=(N_c^2-1)/2N_c$, and $C_B=(N_c+1)/2N_c$. 
Explicit forms of the scalar coefficients $D_i$ defined in Eq.~(\ref{perturb}) 
are listed in Appendix~A. When Eq.~(\ref{perturb}) is expanded in terms of 
a power series in $1/\epsilon$, one obtains
\begin{eqnarray}\label{results1}
P_3(\omega)&=&-\frac{\langle\bar qq\rangle}{2\pi^2}\omega^2
  \Bigg[\left(\frac{-\mu}{2\omega}\right)^{2\epsilon}
  \Bigg(\frac1{\epsilon}+2\Bigg)\nonumber\\&&
  +\frac{\as}{12\pi}\left(\frac{-\mu}{2\omega}\right)^{4\epsilon}
  \Bigg(\frac1{\epsilon^2}(2n^2-8n+7+2(n-2)s)\nonumber\\&&\qquad\qquad
  +\frac1\epsilon(12n^2-44n+51+(12n-22)s+8\zeta(2))\nonumber\\&&\qquad\qquad
  +56n^2-200n+260+(56n-100)s\nonumber\\&&\qquad\qquad
  +(18n^2-72n+87+18s(n-2))\zeta(2)-32\zeta(3)\Bigg)\Bigg],\qquad
\end{eqnarray}
where we have now substituted explicit values for the colour factors with 
$N_c=3$. The spectral density $\rho_3(\omega)$ is given by the absorptive 
part of $P_3(\omega)$. In renormalizing the spectral density 
$\rho_3(\omega)$ one has to take into account both the renormalization 
factor of the baryon currents~\cite{GrYa} and the renormalization factor 
of the quark condensate,
\begin{equation}\label{rho3ren}
\rho_3(\omega)=Z_{J_1}Z_{J_2}Z_{\bar qq}^{-1}\rho_3^{\rm ren}(\omega)
\end{equation}
with~\cite{GKY1,BaBa}
\begin{eqnarray}
Z_{\bar qq}&=&1+3\frac{\as C_F}{4\pi\epsilon},\nonumber\\
Z_{J_1}&=&1+\frac{\as C_B}{4\pi\epsilon}(n^2-4n+6)\quad\mbox{and}\nonumber\\
Z_{J_2}&=&1+\frac{\as C_B}{4\pi\epsilon}((n-2+s)^2+2).
\end{eqnarray}
After multiplication with the $Z$-factors the leading 
$1/\epsilon$-contribution in $\rho_3(\omega)$ is cancelled. The 
renormalized spectral density is given by
\begin{equation}\label{rhoren}
\rho_3^{\rm ren}(\omega)
  =\rho_3^{\rm Born}(\omega)\Bigg[1+\frac{\as}{4\pi}r(\omega/\mu)\Bigg],
\end{equation}
where
\begin{equation}\label{rho}
  \rho_3^{\rm Born}(\omega)=-\frac{\langle\bar qq \rangle^{\rm ren}}{\pi^2}
  \omega^2\quad\mbox{and}\quad
  r(\omega/\mu)=r_1\ln\left(\frac\mu{2\omega}\right)+r_2
\end{equation}
with
\begin{eqnarray}\label{rhocoeff}
  r_1&:=&\frac43(2n^2-8n+7+2(n-2)s)\quad\mbox{and}\nonumber\\
  r_2&:=&\frac23(8n^2-28n+37+8ns-14s+8\zeta(2)).
\end{eqnarray}
Note that the coefficient $r_1$ of the logarithmic term in Eq.~(\ref{rho}) 
coincides with the sum of the one-loop anomalous dimension of $J_1$ and 
$J_2$ minus the anomalous dimension of the quark condensate. The reason 
is that the same coefficient is involved in the cancellations of the 
$1/\epsilon$-pole in Eq.~(\ref{rho3ren}).

Explicit values for the correction to the spectral density for the cases of 
the $\Lambda_Q$- and $\Sigma_Q$-type ground state baryons are given by
\begin{equation}
r_\Lambda(\omega/\mu)=4\ln\left(\frac\mu{2\omega}\right)
  +\frac23(23+8\zeta(2))
\end{equation}
and
\begin{equation}
r_\Sigma(\omega/\mu)=-\frac43\ln\left(\frac\mu{2\omega}\right)
  +\frac23(11+8\zeta(2)).
\end{equation}
The radiative corrections can be seen to amount to about $40-60\%$ at
the renormalization scale $\mu=1\GeV$. Due to the hermiticity of the 
current correlator $\Pi_{ij}$ the coefficients $r_1$ and $r_2$ do not 
depend on which of the two non-diagonal current products $J_1\bar J_2$ or 
$J_2\bar J_1$ are taken.

\subsection{Non-diagonal sum rules}
As usual we construct QCD sum rules by invoking parton-hadron duality,
i.e.\ we equate the theoretical contribution to the scalar correlation 
function $P(\omega)$ given in Eq.~(\ref{OPE}) with the dispersion integral 
over the contributions of hadron states. These consist of the lowest lying 
ground state with bound state energy $\bar\Lambda$ plus the excited states 
and the continuum contributions. To leading order in $1/m_Q$ the bound 
state energy of the ground state is defined by 
\begin{equation}
m_{\rm baryon}=m_Q+\bar\Lambda,
\end{equation}
where $m_Q$ is the pole mass of the heavy quark. We assume that the 
contribution of the excited states and the continuum contribution sets in 
above some effective threshold energy $E_C$ and can be approximated by the 
OPE expression~\cite{Vain}. For the hadron-side contribution $\rho_{\rm HS}$ 
to the spectral density we thus write 
\begin{equation}
\rho_{\rm HS}(\omega)=\rho_{\rm GS}(\omega)+\rho_{\rm cont}(\omega),
\end{equation}
where the contribution of the lowest-lying ground state baryon is denoted 
by $\rho_{\rm GS}$ and is given by
\begin{eqnarray}\label{rhoground}
\rho_{\rm GS}(\omega)=\frac12F_1F_2\delta(\omega-\bar\Lambda).
\end{eqnarray}
The residues $F_i$ ($i=1,2$) appearing in Eq.~(\ref{rhoground}) are defined 
by the matrix elements of the heavy baryon currents according to
\begin{equation}\label{residue}
\langle 0|J_i|\Lambda_Q\rangle=F_{i\Lambda}u,\qquad
\langle 0|J_i|\Sigma_Q\rangle=F_{i\Sigma}u\quad\mbox{and}\quad
\langle 0|J_i^\nu|\Sigma^*_Q\rangle=\frac1{\sqrt 3}F_{i\Sigma^*}u^\nu,
\end{equation}
where $u$ and $u^\nu$ are the usual spin~$1/2$ and spin~$3/2$ spinors.
Note that $F_{i\Sigma^*}$ coincides with $F_{i\Sigma}$ to lowest 
order of the heavy quark mass expansion which we are working in.

As is usual we assume hadron-parton duality for the contributions of 
excited states and the continuum contributions. As mentioned before we 
subsum these contributions by defining an effective energy threshold $E_C$ 
and write $\rho_{\rm cont}(\omega)=\theta(\omega-E_C)\rho(\omega)$, where 
$\rho$ is the result of the OPE calculations given in Eqs.~(\ref{rho3rho5}) 
and~(\ref{rhoren}). With these assumptions we arrive at the sum rule
\begin{equation}
P_{\rm OPE}(\omega)=\frac{\frac12F_1F_2}{\bar\Lambda-\omega-i0}
  +\int_{E_C}^\infty\frac{\rho(\omega')d\omega'}{\omega'-\omega-i0}
\end{equation}
or
\begin{equation}\label{presumrule}
\frac{\frac12F_1F_2}{\bar\Lambda-\omega-i0}=\int_0^{E_C}
  \frac{\rho(\omega')d\omega'}{\omega'-\omega-i0}
  +P_{\rm PC}(\omega).
\end{equation}
The polynomial contribution $P_{\rm PC}(\omega)$ is defined as the Fourier 
transform of that part of the correlator function $P(t)$ which contains 
non-negative powers $(t^2)^n$ ($n\geq 0$). Finally we apply the Borel 
transformation 
\begin{equation}
\hat B_T=\lim\frac{\omega^n}{\Gamma(n)}\left(-\frac{d}{d\omega}\right)^n
  \qquad n,-\omega\to\infty\quad(T=-\omega/n\hbox{\rm\ fixed}) 
\end{equation}
to the sum rule in Eq.~(\ref{presumrule}). Using  
$\hat B_T(1/(\omega-\omega'))=\exp(-\omega'/T)/T$ we obtain the Borel sum rule
\begin{equation}\label{sumrule}
\frac12F_1(\mu)F_2(\mu)e^{-\bar\Lambda/T}
  =\int_0^{E_C}\rho(\omega',\mu)e^{-\omega'/T}d\omega'
  +\hat BP_{\rm PC}(T)=:K(E_C,T,\mu),
\end{equation}
where we have reintroduced the $\mu$-dependence of the spectral density 
which in turn gives rise to a $\mu$-dependence of the residue. The 
Borel-transformed polynomial contribution $\hat BP_{\rm PC}(T)$ can be 
obtained directly from $P_{\rm PC}(t)$ by the substitution $t\rightarrow-i/T$ 
(see the discussion in~\cite{GrYa}). Note that the bound state energy 
$\bar\Lambda$ can be obtained from the sum rule in Eq.~(\ref{sumrule}) by 
taking the logarithmic derivative with respect to the inverse Borel parameter 
according to
\begin{equation}
\bar\Lambda=-\frac{d\ln(K(E_C,T,\mu))}{dT^{-1}}.
\end{equation}

Before turning to the numerical analysis of the non-diagonal sum rules 
we want to briefly comment on the scale dependence of the residues. The 
numerical values given below are taken at the specific normalization point 
$\mu=1\GeV$, while the general dependence on the scale $\mu$ is controlled 
by the renormalization group equation. At one-loop order the product of the 
residues at one renormalization point $\mu_2$ can be expressed by the 
product at another renormalization point $\mu_1$ via
\begin{equation}
F_1(\mu_2)F_2(\mu_2)=F_1(\mu_1)F_2(\mu_1)U(\mu_1,\mu_2)
\end{equation}
with
\begin{equation}\label{evolution1}
  U(\mu_1,\mu_2)=
\left(\frac{\as(\mu)}{\as(\mu_0)}\right)^{\gamma_1/\beta_1},
\end{equation}
where $U(\mu_1,\mu_2)$ is the evolution function which takes one from the
scale $\mu_2$ to the scale $\mu_1$. The coefficients $\beta_1=11-2/3N_c$ 
and $\gamma_1$ are the usual first-order terms in the expansion of the QCD 
$\beta$-function and the anomalous dimension 
$\gamma_P=\gamma_{J_1}+\gamma_{J_2}-\gamma_{\bar qq}$ of the non-diagonal 
scalar correlator function $P$. The two-loop extension of 
Eq.~(\ref{evolution1}) is given by
\begin{eqnarray}\label{evolution2}
U(\mu_1,\mu_2)&=&\exp\left(\int^{\as(\mu_1)}_{\as(\mu_2)}
  \frac{d\alpha}{\alpha}\frac{\gamma(\alpha)}{\beta(\alpha)}\right)
  \nonumber\\&=&\left(\frac{\as(\mu_1)}{\as(\mu_2)}
  \right)^{\gamma_1/\beta_1}\left(1+\frac{\as(\mu_1)-\as(\mu_2)}
  {4\pi}\frac{\gamma_1}{\beta_1}\left(\frac{\gamma_2}{\gamma_1}
  -\frac{\beta_2}{\beta_1}\right)\right),
\end{eqnarray}
where $\beta_n$ is the $n$-th order term in the $\beta$-function expansion 
and $\gamma_n$ denotes the anomalous dimension of the non-diagonal scalar 
correlator function $P$ at $n$-th loop order. The two-loop evolution 
function in Eq.~(\ref{evolution2}) is obtained as a solution to the 
renormalization group equation including next-to-leading order perturbative 
terms in $\as$ (see also the discussion in~\cite{BrGr1,BaBa,Bardeen,JiMu}).

It is evident that we can only extract the value of the product of residues 
$F_1F_2$ from our sum rule analysis. In order to make further progress, we 
adopt the working hypothesis that the residues of the two current options 
in each case are equal. This assumption is corraborated by the results of 
the diagonal sum rule analysis~\cite{GKY1}. This means we replace $F_1F_2$ 
by $F^2$ in the above formulae when performing the numerical analysis. We 
note, however, that the currents $J_1$ and $J_2$ have different anomalous 
dimensions and therefore $F_1$ and $F_2$ do not coincide at another 
renormalization scale $\mu_2$ even if they coincide at the scale $\mu_1$. 
Returning to the sum rule in Eq.~(\ref{sumrule}), one then has
\begin{eqnarray}
\frac12F^2(\mu)e^{-\bar\Lambda/T}
  \!\!&=&\!\!\frac{2N_c!}{\pi^4}\Bigg[E_Q^3T^3\left(
  f_2(x_C)+\frac{\as}{4\pi}\left(\left(\ln\Big(\frac\mu{2T}\Big)
  f_2(x_C)-f_2^l(x_C)\right)r_1+f_2(x_C)r_2\right)\right)\nonumber\\&&
  -E_Q^3E_0^2T\left(1-\frac c2\right)f_0(x_C)
  +\frac23\left(1-\frac c2\right)\frac{E_Q^3E_G^4}T
  +\frac{\as C_F}{36\pi}\frac{E_Q^9}{T^3}\Bigg],\qquad
\end{eqnarray}
where the $(n,s)$-dependent coefficient functions $r_1$ and $r_2$ are 
defined in Eq.~(\ref{rhocoeff}) and the functions $f_n$ and $f_n^l$ are 
given by
\begin{eqnarray}
f_n(x)&:=&\int_0^x\frac{x'^n}{n!}e^{-x'}dx'
  \ =\ 1-e^{-x}\sum_{m=0}^n\frac{x^m}{m!},\nonumber\\
f_n^l(x)&:=&\int_0^x\frac{x'^n}{n!}\ln x'e^{-x'}dx'.
\end{eqnarray}
In order to simplify the notation we have introduced the abbreviations
\begin{equation}
x_C:=\frac{E_C}T,\quad E_0:=\frac{m_0}4,\quad
  (E_Q)^3:=-\frac{\pi^2}{2N_c}\langle\bar qq\rangle\quad\mbox{and}\quad
  (E_G)^4:=\frac{\pi\as\langle G^2\rangle}{32N_c(N_c-1)}.\qquad
\end{equation}

\subsection{Mixed sum rules of constituent type}
As mentioned before, we shall not investigate the most general 
case of mixed sum rules but specify to the linear combination of currents 
$J=(J_1+J_2)/2$ in the sum rules. The light-side Dirac structure of the 
currents can then seen to appear in the form $\frac12(1+\slv)\Gamma$, 
i.e.\ one has the projector factor $P_+=(1+\slv)/2$ which projects on the 
large components of the light quark fields. In the rest system of the heavy 
baryon, where $v_\mu=(1;0,0,0)$, this is manifest since then 
$P_+=\left({1\ 0\atop 0\ 0}\right)$. We refer to this particular linear 
combination of currents as the constituent type current. This linear 
combination of currents is expected to have maximum overlap with the heavy 
ground state baryons in the constituent quark model, i.e.\ where the light 
diquark state in the heavy baryon is taken to be composed of on-shell light 
quarks. We mention that the constituent quark model picture emerges in the 
large $N_c$-limit~\cite{Witten}. The tools needed for the sum rule analysis 
of constituent type heavy baryons have been assembled in this paper and 
in~\cite{GKY}. The results of the constituent analysis are presented in the 
next section together with the results of the analysis of the diagonal and 
non-diagonal sum rules.

\section{Numerical analysis}

Having the neccessary formulae at hand we next describe our numerical 
analysis of the sum rules and specify our choice of the relevant input 
parameters. We use the following numerical input values for the condensate 
contributions~\cite{Vain,Narison} 
\begin{eqnarray}\label{condensates}
\langle\bar qq\rangle&=&-(0.23\GeV)^3\quad \mbox{(quark condensate)},\\
\as\langle G^2\rangle&=&0.04\GeV^4 \quad \mbox{(gluon condensate)},\quad
  \mbox{and}\nonumber\\
\gs\langle\bar q\sigma_{\mu\nu}G^{\mu\nu}q\rangle&=&m_0^2\langle\bar qq\rangle
  \quad\mbox{with}\quad m^2_0=0.8\GeV^2\quad
  \mbox{(mixed quark-gluon condensate)}.\nonumber
\end{eqnarray}
There are in general two strategies for the numerical analysis of the QCD 
sum rules. The first strategy fixes the bound state energy $\bar\Lambda$ 
from the outset by choosing a specific value for the pole mass of the heavy 
quark and then extracts a value for the residue $F$. In order to obtain 
information from the sum rules which is independent of specific input 
values, we adopt a second strategy, namely to determine both $\bar\Lambda$ 
and $F$ by finding simultaneous stability values for them with respect to 
the Borel parameter $T$.

The first step in carrying out the numerical analysis of the sum rules 
is to find a sum rule ``window'' for the allowed values of the Borel 
parameter $T$. The parameter range of $T$ is constrained by two different 
physical requirements. The first is that the convergence of the OPE 
expansion must be secured. We therefore demand that the subleading term 
in the OPE does not contribute more than $30\%$ of the leading order term. 
This gives a lower limit for the Borel parameter. The upper limit is 
determined by the requirement that the contributions from the excited 
states plus the physical continuum (even after Borel transformation) 
should not exceed the bound state contribution. This requirement is 
neccessary in order to guarantee that the sum rules are as independent as 
possible of the model-dependent assumptions concerning the profile of the 
theoretical spectral density, i.e.\ the model of the continuum.

The lower limit of $E_C$ is given by the requirement that the indicated 
window should be kept open. For the rest, $E_C$ is a free floating variable 
which is only limited by the stability requirements on $\bar\Lambda$ and $F$.

\subsection{Diagonal sum rules}
Let us briefly recapitulate the results of the numerical analysis of the 
diagonal sum rules presented in~\cite{GKY1}. The above two requirements 
limit the allowed range for the Borel parameter to $250\MeV<T<400\MeV$. The 
analysis proceeded in two steps. First we analyzed the uncorrected sum 
rules varying both the continuum threshold and the bound state energy. The 
criterion for the best choice of these two energies is the stability of the 
sum rules with regard to the variation of the Borel parameter $T$. In the 
second step we included the radiative corrections and again varied both the 
continuum threshold and the bound state energy to obtain the best sum rules 
stability. The ratio of the continuum contribution and ground state 
contribution depends strongly on the Borel parameter $T$ and the continuum 
threshold energy $E_C$. Looking e.g.\ at the sum rule analysis for the 
$\Lambda_Q$-type baryons with QCD corrections, the continuum contribution is 
about 80\% of the ground state contribution for $E_C=1.1\GeV$ and 
$T=250\MeV$ and then increases with~$T$.

Because of the new specification for the sum rule window we have repeated 
the diagonal sum rule analysis of~\cite{GKY1} allowing for slightly different 
values of $\bar\Lambda$ and $F$. The outcome of the numerical analysis is 
practically unaltered. The values for the $\Lambda_Q$-type state can be read 
off from Fig.~3. Fig~3(a) shows the dependence of the bound state energy 
$\bar\Lambda$ on the Borel parameter $T$ and Fig.~3(b) shows the dependence 
of the residue on $T$, both for the leading order sum rule. Fig.~3(c) and 
Fig.~3(d) show the same dependencies for the radiatively corrected sum 
rules. The same analysis is repeated for the $\Sigma_Q$-type states in 
Fig.~4. The results of the numerical analysis both without and with 
radiative corrections are given in Table~\ref{tab2}.
\begin{table}\begin{center}
\begin{tabular}{|c||c|c|c|}
Baryon type state&$E_C$&$\bar\Lambda$&$F$\\\hline
$\Lambda_Q$ (L.O.)
  &$1.2\pm 0.1\GeV$&$0.77\pm 0.05\GeV$&$0.022\pm 0.001\GeV^3$\\
$\Lambda_Q$ (N.L.O.)
  &$1.1\pm 0.1\GeV$&$0.77\pm 0.05\GeV$&$0.027\pm 0.002\GeV^3$\\
$\Sigma_Q$ (L.O.)
  &$1.4\pm 0.1\GeV$&$0.96\pm 0.05\GeV$&$0.031\pm 0.002\GeV^3$\\
$\Sigma_Q$ (N.L.O.)
  &$1.3\pm 0.1\GeV$&$0.94\pm 0.05\GeV$&$0.038\pm 0.003\GeV^3$\\
\hline\end{tabular}
\caption{\label{tab2}Results of the diagonal sum rule analysis for the 
continuum threshold parameter $E_C$, the bound state energy $\bar\Lambda$, 
and the residuum $F$ for $\Lambda_Q$-type and $\Sigma_Q$-type currents, 
analyzed to leading order (L.O.) as well as next-to-leading order (N.L.O.)}
\end{center}\end{table}

\subsection{Non-diagonal sum rules}
In the case of the non-diagonal sum rules, the ``window'' of permissible 
values for the Borel parameter is wider than in the diagonal case and it 
is given by $250\MeV<T<600\MeV$. Proceeding in the same manner as in the 
case of the diagonal sum rule analysis, we obtain best stability values 
when varying $T$. The values for the $\Lambda_Q$-type state can be read off 
from Fig.~5, and the values for the $\Sigma_Q$-type state can be obtained 
from Fig.~6. For the $\Lambda_Q$-type baryons the stability appears at 
values of $T$ and $E_C$ where the continuum contribution is about $100\%$. 
If we try to decrease this contribution relatively by increasing $E_C$, 
the stability in $T$ becomes worse. As can be seen from Fig.~5(a), for e.g.\ 
$E_C=1.3\GeV$ the contribution of the continuum is less than $40\%$ on the 
left hand side, but stability is lost. These considerations show that the 
relative error of our estimate can be taken to be approximately $10\%$. The 
situation for the $\Sigma_Q$-type baryons is much better. For example, for 
the radiatively corrected sum rules the radio of the continuum and the 
ground state contribution is $50\%$ for the central value $E_C=1.2\GeV$ and 
$30\%$ for $E_C=1.5\GeV$ at the left end of the allowed range for the Borel 
parameter $T$.

The numerical results are given in Table~\ref{tab3}. Assuming relative 
errors of $10\%$ for the bound state energy and $20\%$ for the residue, the 
obtained values are in agreement with the results of the analysis of the 
diagonal sum rules, where the values for the $\Sigma_Q$-type baryon are the 
more reliable one.
\begin{table}\begin{center}
\begin{tabular}{|c||c|c|c|}
Baryon type state&$E_C$&$\bar\Lambda$&$F$\\\hline
$\Lambda_Q$ (L.O.)
  &$1.0\pm 0.10\GeV$&$0.75\pm 0.10\GeV$&$0.024\pm 0.002\GeV^3$\\
$\Lambda_Q$ (N.L.O.)
  &$1.0\pm 0.10\GeV$&$0.72\pm 0.10\GeV$&$0.032\pm 0.003\GeV^3$\\
$\Sigma_Q$ (L.O.)
  &$1.5\pm 0.10\GeV$&$1.16\pm 0.10\GeV$&$0.045\pm 0.003\GeV^3$\\
$\Sigma_Q$ (N.L.O.)
  &$1.2\pm 0.10\GeV$&$0.94\pm 0.10\GeV$&$0.039\pm 0.004\GeV^3$\\
\hline\end{tabular}
\caption{\label{tab3}Results of the non-diagonal sum rule analysis for the 
continuum threshold parameter $E_C$, the bound state energy $\bar\Lambda$, 
and the residuum $F$ for $\Lambda_Q$-type and $\Sigma_Q$-type currents, 
analyzed to leading order (L.O.) as well as next-to-leading order (N.L.O.)}
\end{center}\end{table}

\subsection{Constituent type mixed sum rules}
The use of a constituent type interpolating current $J=(J_1+J_2)/2$ 
combines the two sum rule formulas for the diagonal and the non-diagonal 
case, taking one half of each part. The ``window'' of permissible values 
for the Borel parameter $T$ is now given by $300\MeV<T<700\MeV$. In Fig.~7 
we show the results of the sum rule analysis for the $\Lambda_Q$-type 
baryons, and in Fig.~8 we show the results for the $\Sigma_Q$-type baryons.

The numerical results of the analysis are given in Table~\ref{tab4}.
\begin{table}\begin{center}
\begin{tabular}{|c||c|c|c|}
Baryon type state&$E_C$&$\bar\Lambda$&$F$\\\hline
$\Lambda_Q$ (L.O.)
  &$1.1\pm 0.10\GeV$&$0.77\pm 0.10\GeV$&$0.034\pm 0.004\GeV^3$\\
$\Lambda_Q$ (N.L.O.)
  &$1.1\pm 0.10\GeV$&$0.77\pm 0.10\GeV$&$0.032\pm 0.004\GeV^3$\\
$\Sigma_Q$ (L.O.)
  &$1.3\pm 0.10\GeV$&$1.03\pm 0.10\GeV$&$0.045\pm 0.004\GeV^3$\\
$\Sigma_Q$ (N.L.O.)
  &$1.2\pm 0.10\GeV$&$0.94\pm 0.10\GeV$&$0.036\pm 0.004\GeV^3$\\
\hline\end{tabular}
\caption{\label{tab4}Results of the constituent type mixed sum rule analysis 
for the continuum threshold parameter $E_C$, the bound state energy 
$\bar\Lambda$, and the residuum $F$ for $\Lambda_Q$-type and $\Sigma_Q$-type 
currents, analyzed to leading order (L.O.) as well as next-to-leading order 
(N.L.O.)}
\end{center}\end{table}
The constituent type sum rules show an improved stability on the Borel 
parameter $T$ as compared to the non-diagonal sum rules, but the stability 
is not as good as in the diagonal case. Within the assumed errors the 
results are again in agreement with both the diagonal and the non-diagonal 
sum rule analysis.

\subsection{Comparison with experimental values}
Finally we want to compare our results for the bound state energy with the 
existing experimental values for the baryon masses. For such a comparision 
we need to know the pole mass of the heavy quarks which can be extracted 
from the heavy quarkonium and heavy-light mesons~\cite{Vol,Reind,Nar}. The 
quoted value of the bottom quark pole mass varies from 
$m_b=4.55\pm 0.05\GeV$~\cite{Reind} and $m_b=4.67\pm 0.10\GeV$~\cite{Nar}
to $m_b=4.80\pm 0.03\GeV$~\cite{Vol}. Assuming the range 
$4.6\GeV<m_b<4.9\GeV$, the mass $m(\Lambda_b)=5642\pm 50 MeV$ of the baryon 
$\Lambda_b$~\cite{PDG} results in a range 
$740\MeV<\bar\Lambda(\Lambda_b)<1040\MeV$ for the bound state energy. Our 
central value $\bar\Lambda(\Lambda_Q)=760\MeV$ for the bound state energy 
suggests a pole mass of $m_b=4880 MeV$ for the bottom quark.

Taking the experimental results for charm-quark baryons, namely 
$m(\Lambda_c)=2284.9\pm 0.6\MeV$ and 
$m(\Sigma^+_c)=2453.5\pm 0.9\MeV$~\cite{PDG}, our central values 
$\bar\Lambda(\Lambda_Q)=760\MeV$ and $\bar\Lambda(\Sigma_Q)=940\MeV$ 
predict a mean pole mass of $m_c=1520\MeV$ for the charm quark.

\section{Conclusions}
We have considered the operator product expansion of the correlator of two
static heavy baryon currents at small Euclidian distances and determined
the $\as$ radiative corrections to the first and second Wilson coefficient 
in the expansion. Based on the operator product expansion we have 
formulated and analyzed heavy baryon sum rules for the $\Lambda_Q$-type and 
$\Sigma_Q$-type heavy baryons using two different types of interpolating 
fields for the baryons in each case. In this paper we have constructed and 
analyzed the non-diagonal sum rules built from the correlators of two 
different currents including radiative corrections. The non-diagonal sum 
rules bring in some new features such as a more ``normal'' behaviour of the 
spectral density $\rho(\omega)\approx\langle\bar qq\rangle\omega^2$ and 
moderate QCD corrections to the spectral density as compared to the 
diagonal case. We have taken a second look at the diagonal sum rules.

We have also set up and analyzed constituent type heavy baryon sum rules 
where we have used interpolating currents that are expected to have a 
maximum overlap with the heavy baryon's light diquark system in the 
constituent quark model picture. All the three types of sum rules show 
acceptable stability in their dependence on the Borel parameter, where the 
best stability was obtained for the diagonal sum rules. The results of the 
three types of sum rules (diagonal, non-diagonal, constituent type) on the 
bound state energy an the residues of the heavy ground state baryons 
were found to be consistent with each other, where the values obtained for 
the $\Sigma_Q$-type baryons are more reliable than the results for the 
$\Lambda_Q$-type baryons.

\vspace{.5truecm}

\noindent{\large \bf Acknowledgments:}\smallskip\\
This work was partially supported by the BMBF, FRG, under contract 06MZ865, 
and by the Human Capital and Mobility program under contract CHRX-CT94-0579. 
The work of O.I.Y. was supported by the BMBF, FRG, under contract 
057WZ91P(0). We would like to thank A.~Grozin, B.~Tausk and A.~Khodjamirian 
for valuable discussions.

\newpage

\section*{Appendix}
\setcounter{equation}{0}
\def\theequation{A\arabic{equation}}
In this appendix we collect our results on the evaluation of the one-loop 
and two-loop contributions to the non-diagonal correlators of two heavy 
baryon currents. The contributing diagrams are shown in Fig.~2. Introducing 
the abbreviation $E_n=\Gamma(1-\epsilon)^n\Gamma(1+n\epsilon)$ (with 
integer numbers $n=1,2,3,\dots$~) we obtain
\begin{eqnarray}
D_0&=&\frac{4\tilde\Gamma_0E_1}{(D-4)(D-3)},\qquad D_1\ =\ D_2\ =\
  \frac{2(D-2)\tilde\Gamma_0E_2}{(D-4)^2(D-3)(2D-7)},\qquad\\[12pt]
D_3&=&\frac{8(D-2)\tilde\Gamma_0E_1^2}{(D-4)^3(D-3)^2}
  -\frac{4(D-2)(3D-10)\tilde\Gamma_0E_2}{(D-4)^3(D-3)^2(2D-7)},\\[12pt]
D_4&=&\frac{(D-4)\tilde\Gamma_1+\tilde\Gamma_2}{(D-4)^2(D-3)(2D-7)}E_2,\qquad
D_5\ =\ \frac{2(D-2)\tilde\Gamma_0-D\tilde\Gamma_1
  +\tilde\Gamma_2}{(D-4)^2(D-3)(2D-7)}E_2,\qquad\\[12pt]
D_6&=&\frac{2D(D-2)\tilde\Gamma_0E_2}{(D-4)^2(D-3)(2D-7)},\qquad
D_7\ =\ \frac{2(D-2)\tilde\Gamma_0E_2}{(D-4)^2(D-3)(2D-7)},\qquad\\[12pt]
D_8&=&\frac{-4(D-2)\tilde\Gamma_0E_2}{(D-4)^2(D-3)^2(2D-7)}
\end{eqnarray}
where $\tilde\Gamma_0=\spur(\bar\Gamma\slv\Gamma\slv)$,
$\tilde\Gamma_1=\spur(\bar\Gamma\gamma_\mu\Gamma\gamma^\mu)$, and
$\tilde\Gamma_2=\spur(\bar\Gamma\slv\gamma_\mu\gamma_\nu\Gamma
  \gamma^\nu\gamma^\mu\slv)$.

\vspace{1cm}

\centerline{\Large\bf Figure Captions}
\vspace{.5cm}
\newcounter{fig}
\begin{list}{\bf\rm Fig.\ \arabic{fig}:}{\usecounter{fig}
\labelwidth1.6cm\leftmargin2.5cm\labelsep.4cm\itemsep0ex plus.2ex}

\item Radiative corrections to the diagonal correlator. (0) lowest order 
two-loop contribution, (1)--(4) $O(\as)$ three-loop contributions.

\item Radiative corrections to the non-diagonal correlator given by the 
dimension three condensate contribution. (0) lowest order one-loop 
contribution, (1)--(8) $O(\as)$ two-loop contributions.

\item Bound state energy and residue of the $\Lambda_Q$ as functions of the 
Borel parameter $T$ (diagonal case). Plotted are five curves for five 
different values of the threshold energy $E_C$ spaced by $100\MeV$ around 
the central value $E_C=E_C^{\rm best}$. $E_C$ increases from bottom to top.\\
(a) lowest order sum rule results for the bound state energy 
    $\bar\Lambda(\Lambda)$\\
(b) lowest order sum rule results for the residue $F_\Lambda$\\
(c) $O(\as)$ sum rule results for the bound state energy 
    $\bar\Lambda(\Lambda)$\\
    for the current $J_{\Lambda1}$\\
(d) $O(\as)$ sum rule results for the residue $F_\Lambda$
    for the current $J_{\Lambda1}$

\item Bound state energy and residue of the $\Sigma_Q$ as functions of the 
Borel parameter $T$ (diagonal case). Plotted are five curves for five 
different values of the threshold energy $E_C$ spaced by $100\MeV$ around 
the central value $E_C=E_C^{\rm best}$. $E_C$ increases from bottom to top.\\
(a) lowest order sum rule results for the bound state energy 
    $\bar\Lambda(\Sigma)$\\
(b) lowest order sum rule results for the residue $F_\Sigma$\\
(c) $O(\as)$ sum rule results for the bound state energy 
    $\bar\Lambda(\Sigma)$\\
    for the current $J_{\Sigma1}$\\
(d) $O(\as)$ sum rule results for the residue $F_\Sigma$
    for the current $J_{\Sigma1}$

\item Bound state energy and residue of the $\Lambda_Q$ as functions of the 
Borel parameter $T$ (non-diagonal case). Plotted are five curves for five 
different values of the threshold energy $E_C$ spaced by $100\MeV$ around 
the central value $E_C=E_C^{\rm best}$. $E_C$ increases from bottom to top.\\
(a) lowest order sum rule results for the bound state energy 
    $\bar\Lambda(\Lambda)$\\
(b) lowest order sum rule results for the residue $F_\Lambda$\\
(c) $O(\as)$ sum rule results for the bound state energy 
    $\bar\Lambda(\Lambda)$\\
    for the two currents $J_{\Lambda1}$ and $J_{\Lambda2}$\\
(d) $O(\as)$ sum rule results for the residue $F_\Lambda$\\
    for the two currents $J_{\Lambda1}$ and $J_{\Lambda2}$

\item Bound state energy and residue of the $\Sigma_Q$ as functions of the 
Borel parameter $T$ (non-diagonal case). Plotted are five curves for five 
different values of the threshold energy $E_C$ spaced by $100\MeV$ around 
the central value $E_C=E_C^{\rm best}$. $E_C$ increases from bottom to top.\\
(a) lowest order sum rule results for the bound state energy 
    $\bar\Lambda(\Sigma)$\\
(b) lowest order sum rule results for the residue $F_\Sigma$\\
(c) $O(\as)$ sum rule results for the bound state energy 
    $\bar\Lambda(\Sigma)$\\
    for the two currents $J_{\Sigma1}$ and $J_{\Sigma2}$\\
(d) $O(\as)$ sum rule results for the residue $F_\Sigma$\\
    for the two currents $J_{\Sigma1}$ and $J_{\Sigma2}$

\item Bound state energy and residue of the $\Lambda_Q$ as functions of the 
Borel parameter $T$ (constituent type mixed case). Plotted are five curves 
for five different values of the threshold energy $E_C$ spaced by $100\MeV$ 
around the central value $E_C=E_C^{\rm best}$. $E_C$ increases from bottom 
to top.\\
(a) lowest order sum rule results for the bound state energy 
    $\bar\Lambda(\Lambda)$\\
(b) lowest order sum rule results for the residue $F_\Lambda$\\
(c) $O(\as)$ sum rule results for the bound state energy 
    $\bar\Lambda(\Lambda)$\\
    for the two currents $J_{\Lambda1}$ and $J_{\Lambda2}$\\
(d) $O(\as)$ sum rule results for the residue $F_\Lambda$\\
    for the two currents $J_{\Lambda1}$ and $J_{\Lambda2}$

\item Bound state energy and residue of the $\Sigma_Q$ as functions of the 
Borel parameter $T$ (constituent type mixed case). Plotted are five curves 
for five different values of the threshold energy $E_C$ spaced by $100\MeV$ 
around the central value $E_C=E_C^{\rm best}$. $E_C$ increases from bottom 
to top.\\
(a) lowest order sum rule results for the bound state energy 
    $\bar\Lambda(\Sigma)$\\
(b) lowest order sum rule results for the residue $F_\Sigma$\\
(c) $O(\as)$ sum rule results for the bound state energy 
    $\bar\Lambda(\Sigma)$\\
    for the two currents $J_{\Sigma1}$ and $J_{\Sigma2}$\\
(d) $O(\as)$ sum rule results for the residue $F_\Sigma$\\
    for the two currents $J_{\Sigma1}$ and $J_{\Sigma2}$

\end{list}


\begin{thebibliography}{99}
\bibitem{Neub}M.~Neubert, ``Heavy Quark Symmetry'',
  Phys.~Rep.\ {\bf 245} (1994) 259
\bibitem{Vain}M.A.~Shifman, A.I.~Vainstein and V.I.~Zakharov,\\
  Nucl.~Phys.\ {\bf B147} (1979) 385; {\bf B147} (1979) 448
\bibitem{Shur}E.V.~Shuryak, Nucl.~Phys.\ {\bf B198} (1982) 83 
\bibitem{GrYa}A.G.~Grozin and O.I.~Yakovlev,
  Phys.~Lett.\ {\bf 285 B} (1992) 254
\bibitem{BaCha}E.~Bagan, M.~Chabab, H.G.~Dosch and S.~Narison,
  Phys.~Lett.\ {\bf 301 B} (1993) 243
\bibitem{Blok}V.M.~Belyaev and B.Y.~Blok, Z.~Phys.\ {\bf C30} (1986) 151;\\
  B.Y.~Blok and V.L.~Eletzky, Z.~Phys.\ {\bf C30} (1986) 229
\bibitem{BaDo} E.~Bagan, M.~Chabab, H.G.~Dosch and S.~Narison,\\
  Phys.~Lett.\ {\bf 278 B} (1992) 367; {\bf 287 B} (1992) 176
\bibitem{Cher}V.L.~Chernyak and I.R.~Zhitnitsky,
  Nucl.~Phys.\ {\bf B246} (1984) 52;\\
  V.L.~Chernyak, A.A.~Ogloblin and I.R.~Zhitnitsky,
  Z.~Phys.\ {\bf C42} (1989) 569
\bibitem{GKY}S.~Groote, J.G.~K\"orner and O.I.~Yakovlev,
  Phys.~Rev.\ {\bf D54} (1996) 3447
\bibitem{GKY1}S.~Groote, J.G.~K\"orner and O.I.~Yakovlev,
  Phys.~Rev.\ {\bf D55} (1997) 3016
\bibitem{BrGr1}D.~Broadhurst, A.G.~Grozin, Phys.~Lett.\ {\bf 274 B} (1992) 421
\bibitem{BaBa}E.~Bagan, P.~Ball, V.M.~Braun and H.G.~Dosch,
  Phys.~Lett.\ {\bf 278 B} (1992) 457
\bibitem{Bardeen} W.A.~Bardeen, A.J.~Buras, D.W.~Duke and T.~Muta,
  Phys.~Rev.\ {\bf D18} (1978) 3998
\bibitem{JiMu} X.~Ji and M.J.~Musolf, Phys.~Lett.\ {\bf 257 B} (1991) 409
\bibitem{BrGr}D.J.~Broadhurst and A.G.~Grozin,
  Phys.~Lett.\ {\bf 267 B} (1991) 105
\bibitem{Witten}E.~Witten, Nucl.~Phys.\ {\bf B223} (1983) 483;\\
  C.~Carone, H.~Georgi and S.~Osofski, Phys.~Lett.\ {\bf 332 B} (1994) 483;\\
  M.~Luty and J.~March-Russel, Nucl.~Phys. {\bf B246} (1994) 71;\\
  R.F.~Dashen, E.~Jenkins and A.V.~Manohar, Phys.~Rev. {\bf D49} (1994) 4713;\\
  Phys.~Rev.\ {\bf D51} (1995) 3697
\bibitem{Narison} S. Narison, ``QCD Sum Rules'', World Scientific, 
Singapore, 1989 
\bibitem{Vol} M.B.~Voloshin and Y.M.~Zaitsev,
  Sov.~Phys.~Usp.\ {\bf 30} (1987) 553
\bibitem{Reind} L.J.~Reinders, Phys.~Rev.\ {\bf D38} (1988) 947
\bibitem{Nar} S.~Narison, Phys.~Lett.\ {\bf 197 B} (1987) 405
\bibitem{PDG} Particle Data Group, R.M.~Barnett {\it et al},
Phys.~Rev.\ {\bf D50} (1996) 1

\end{thebibliography}
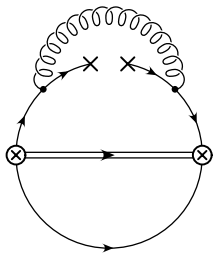
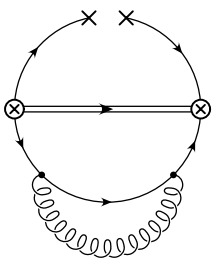
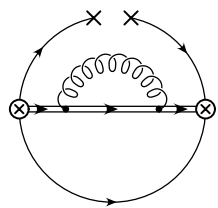
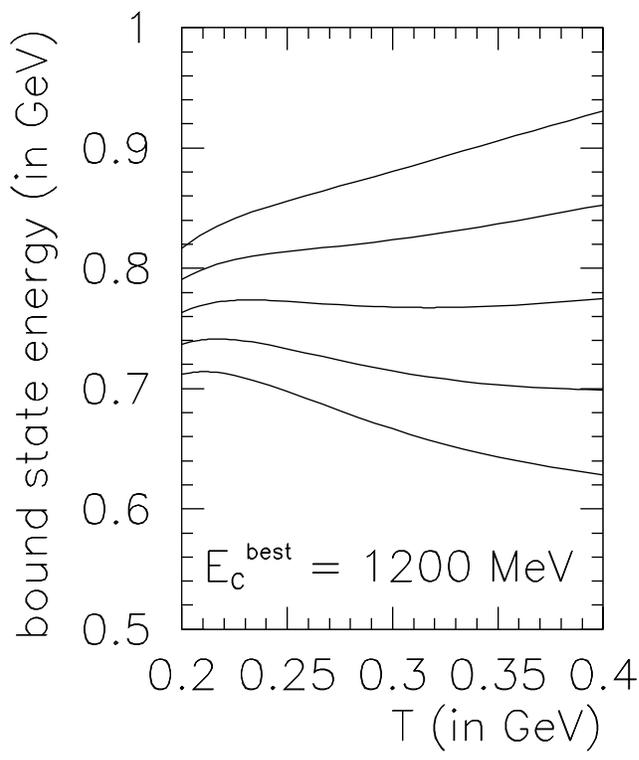
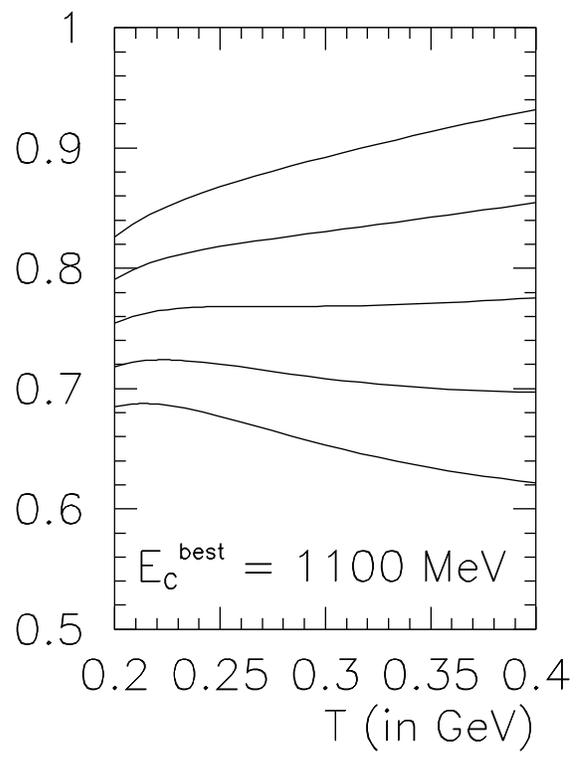
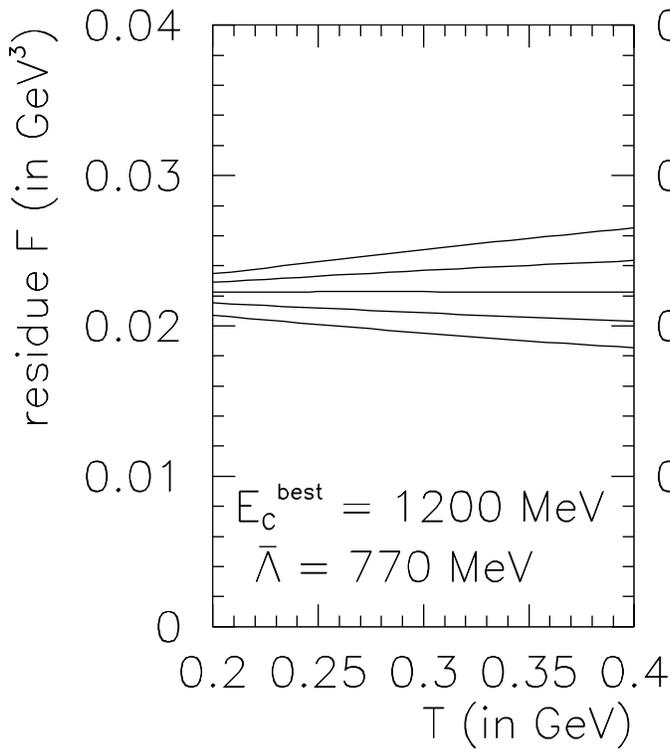
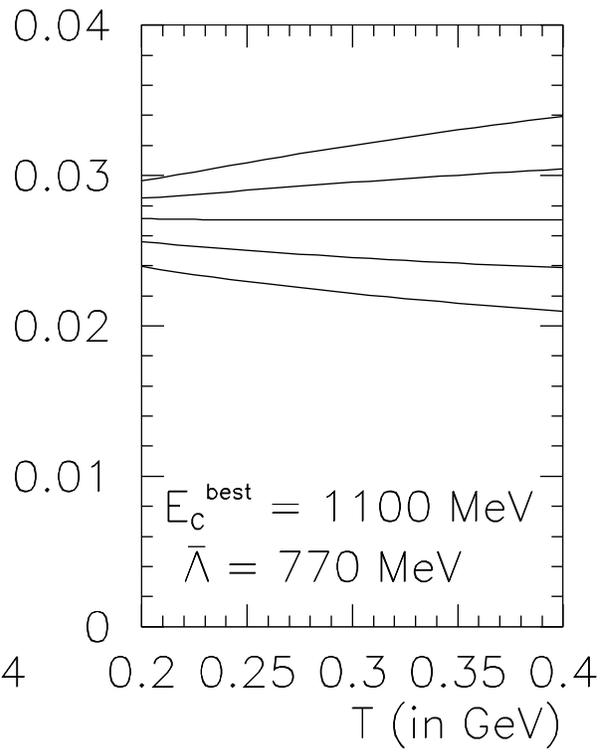
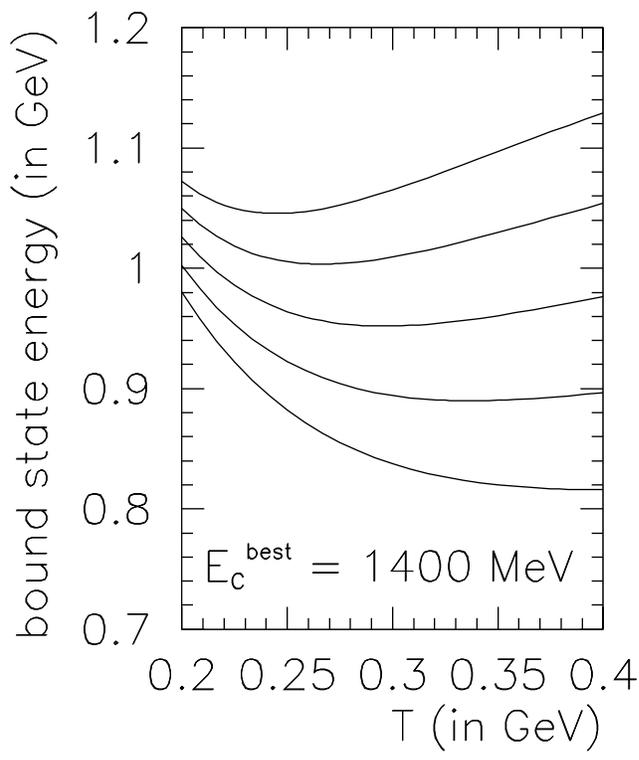
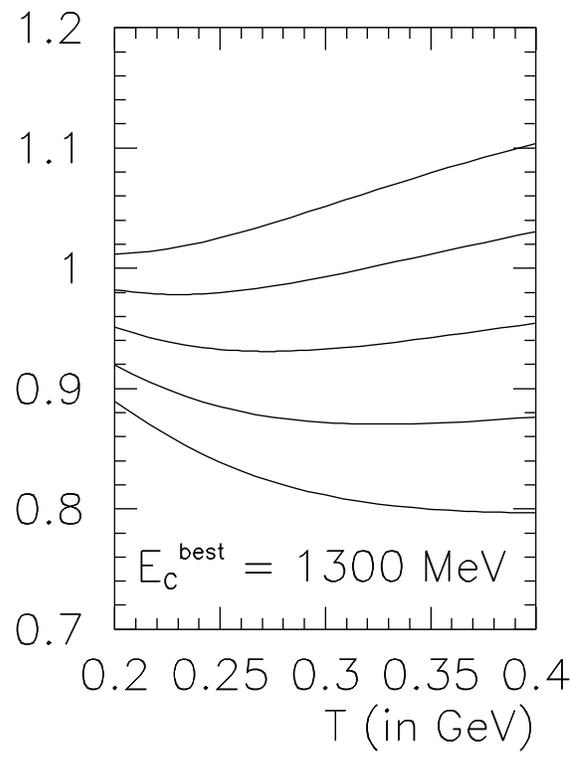
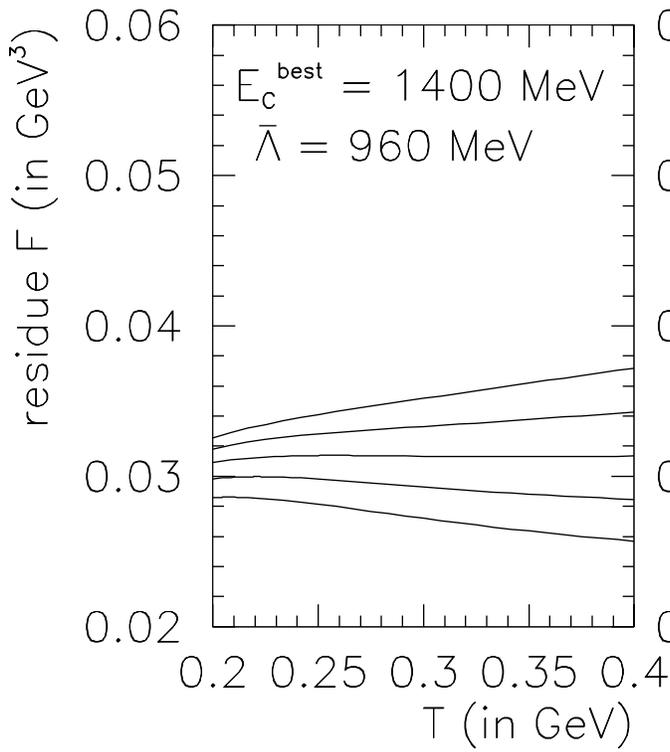
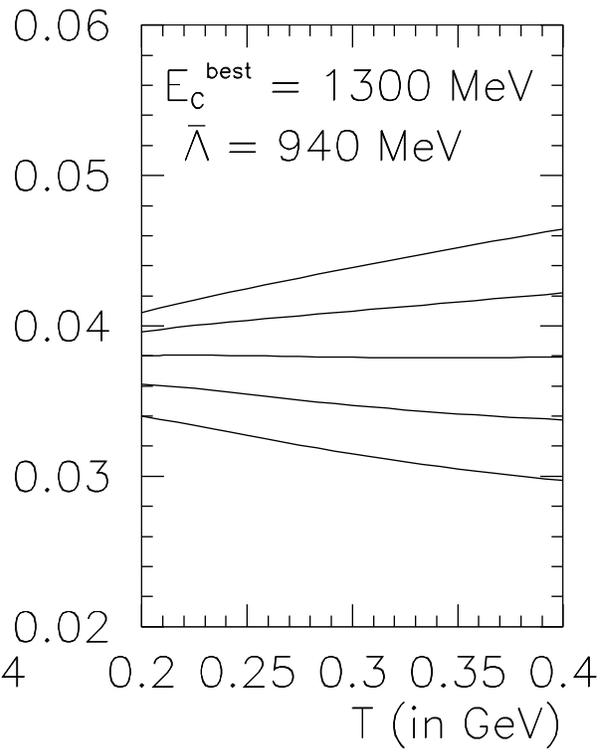
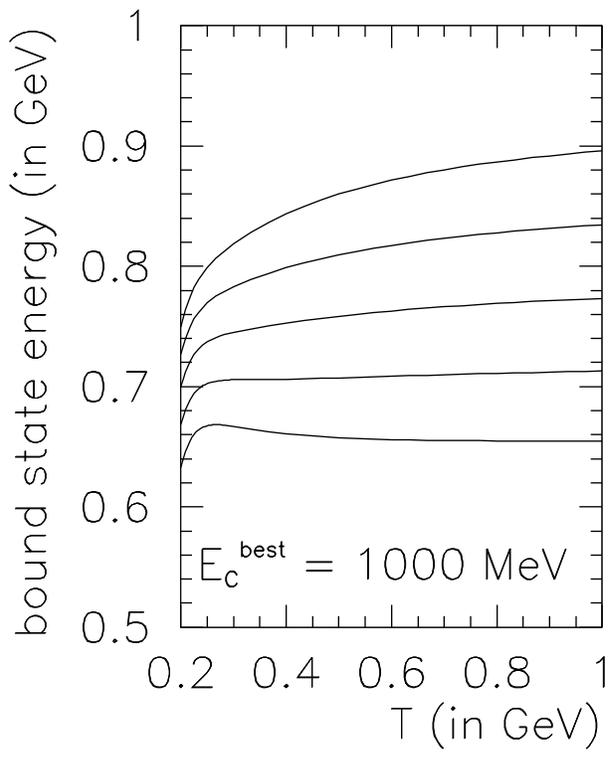
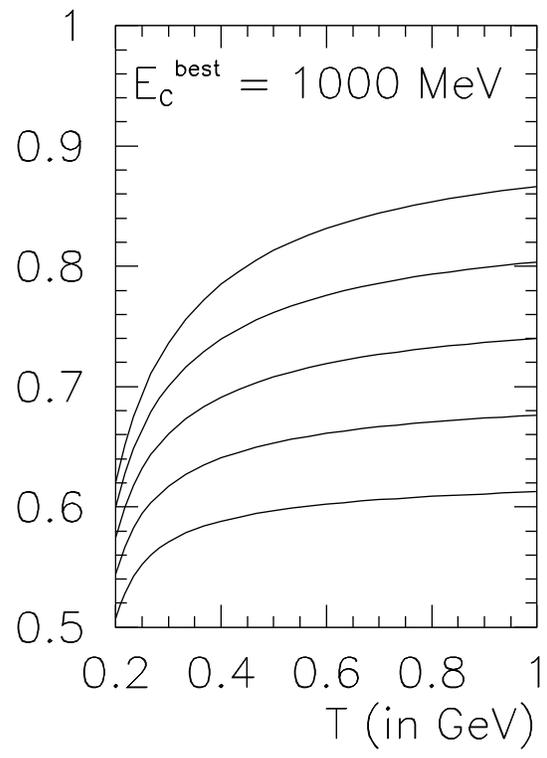
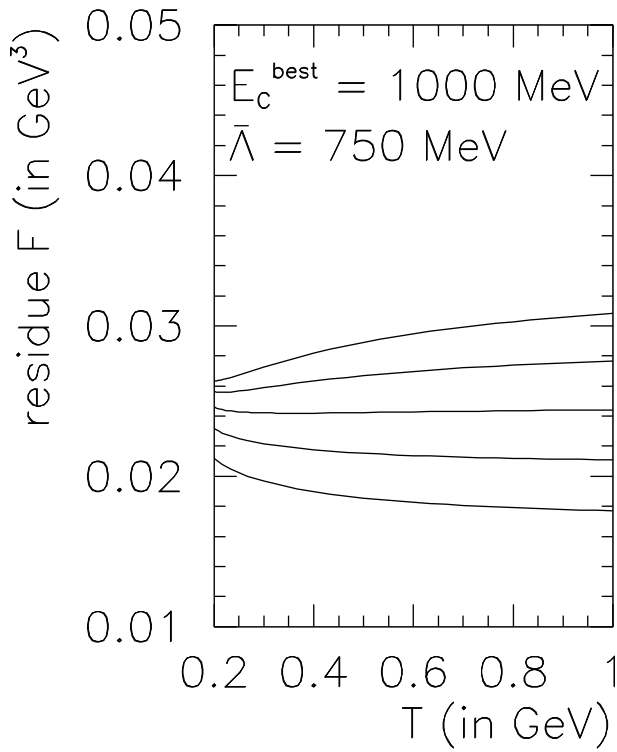
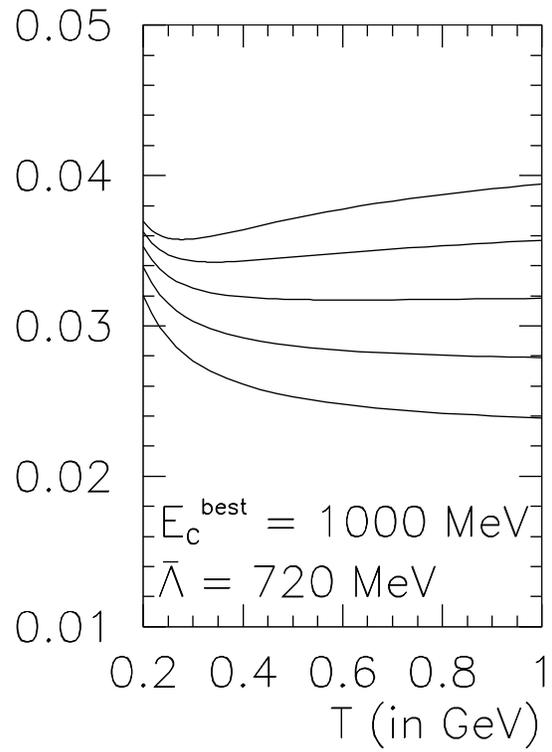
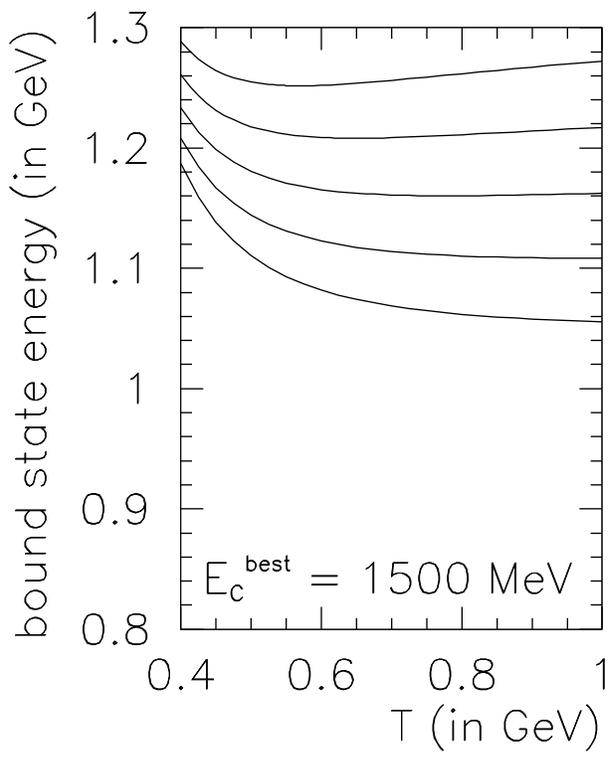
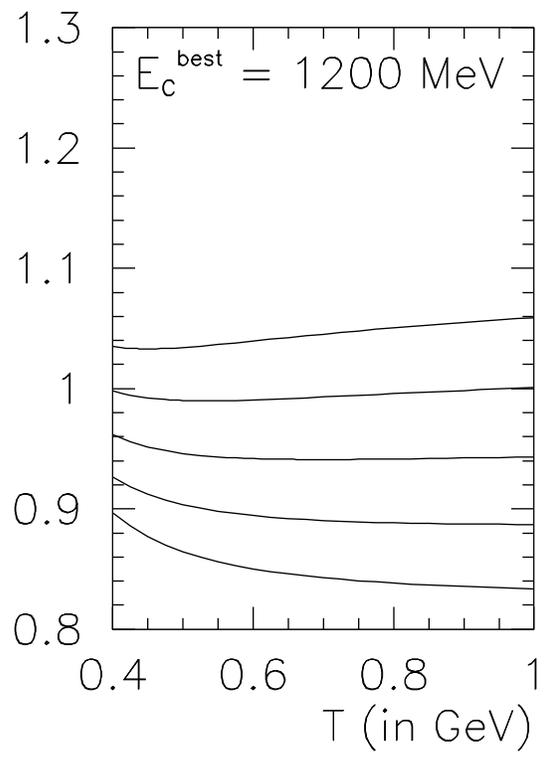
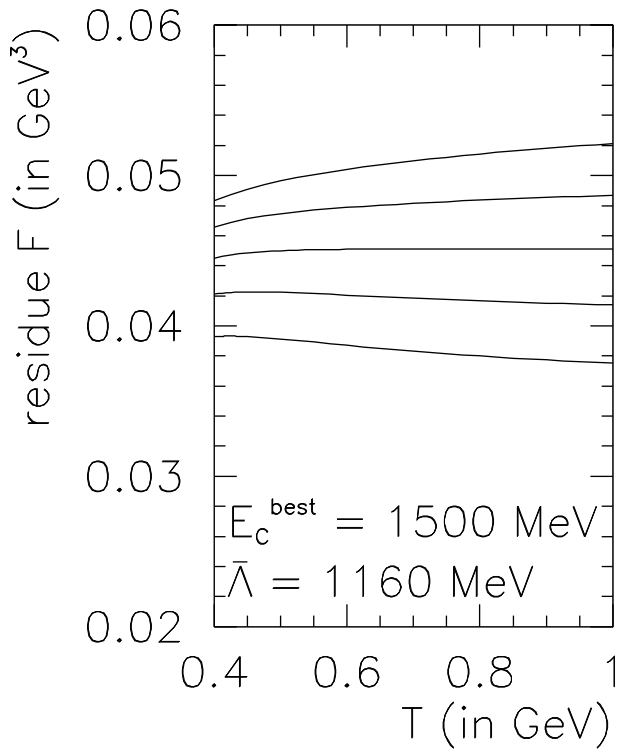
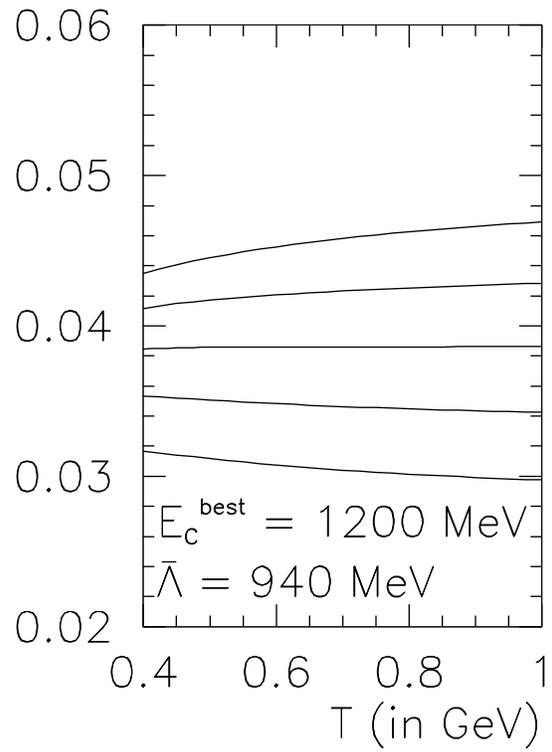
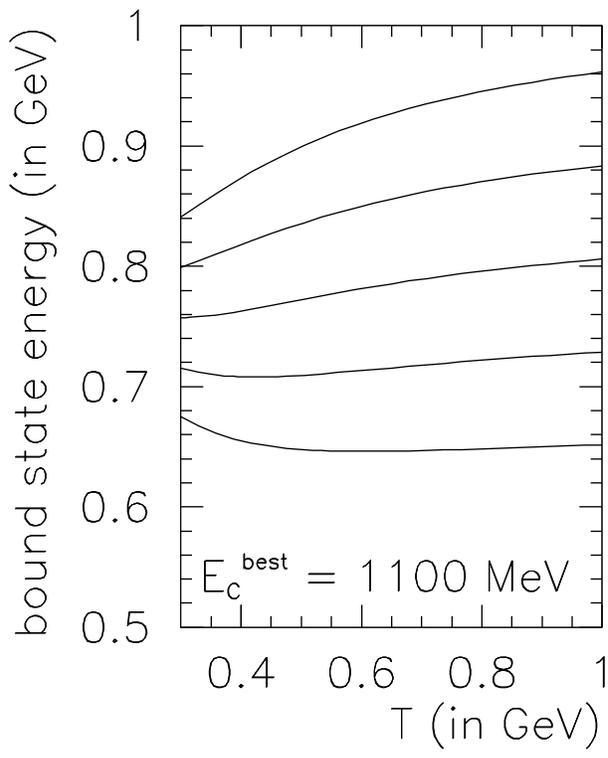
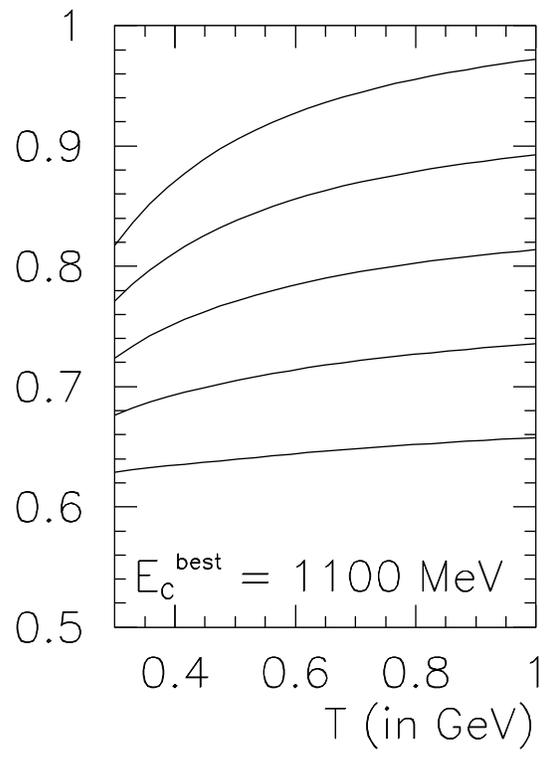
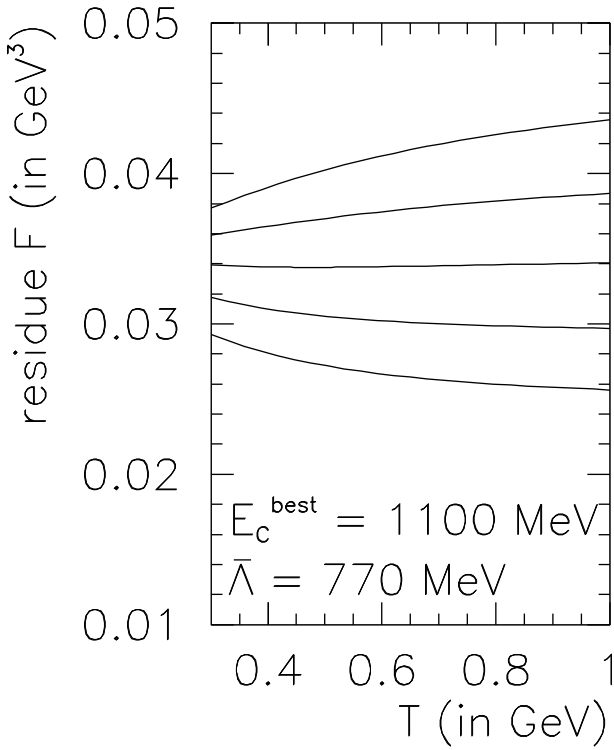
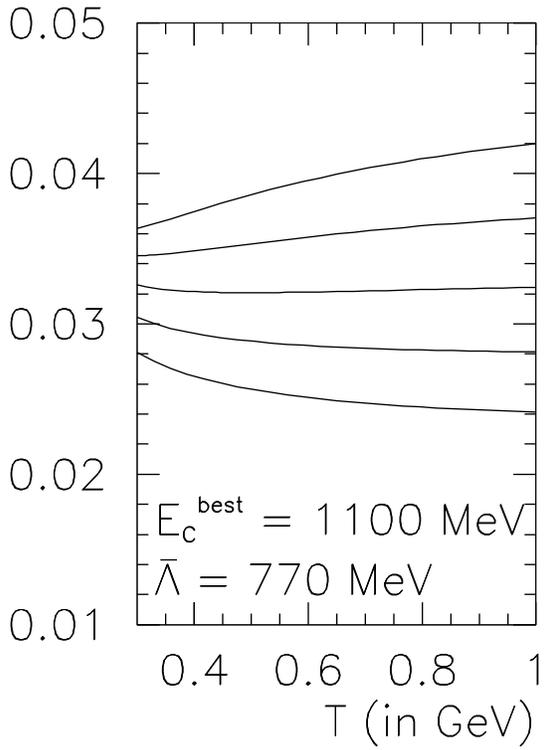
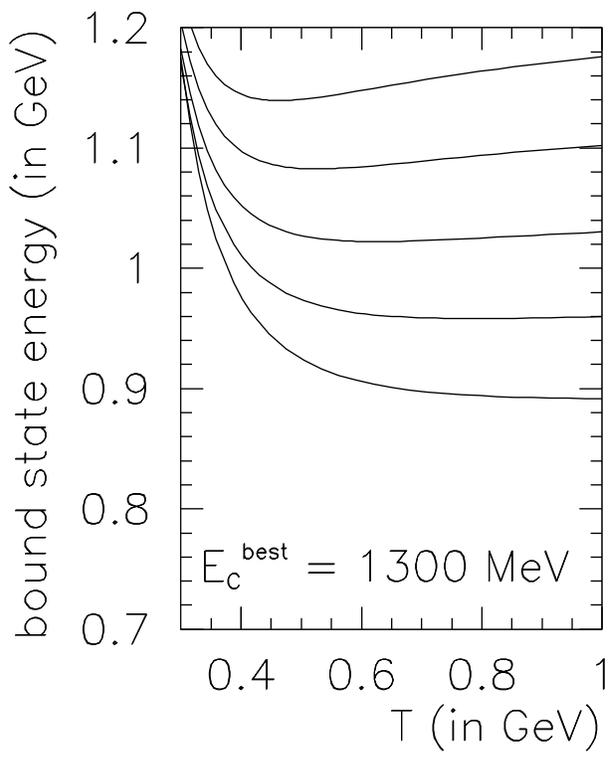
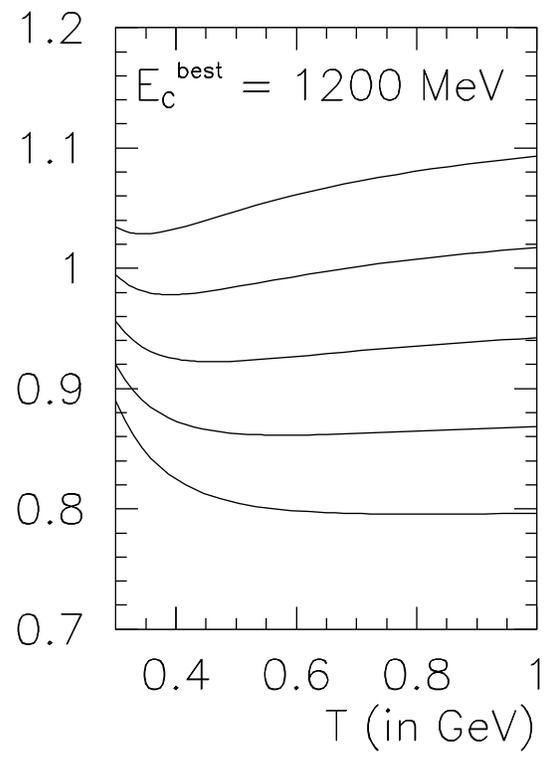
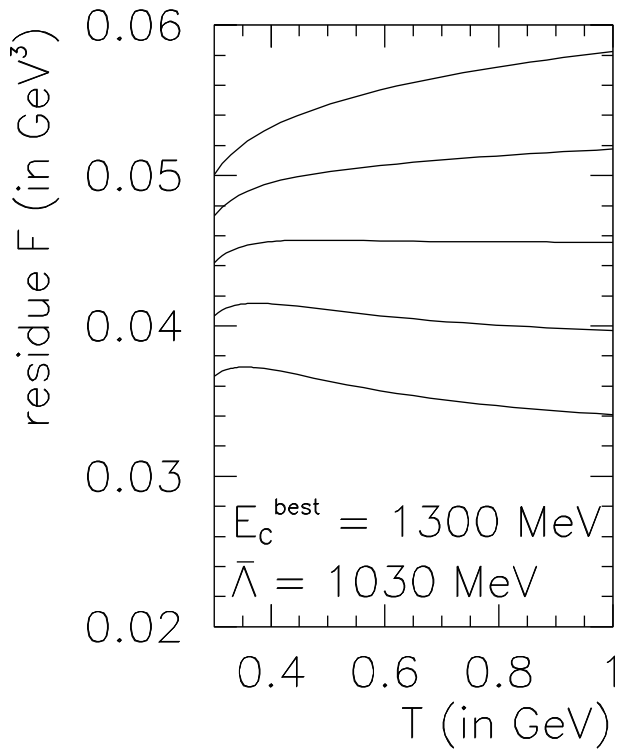
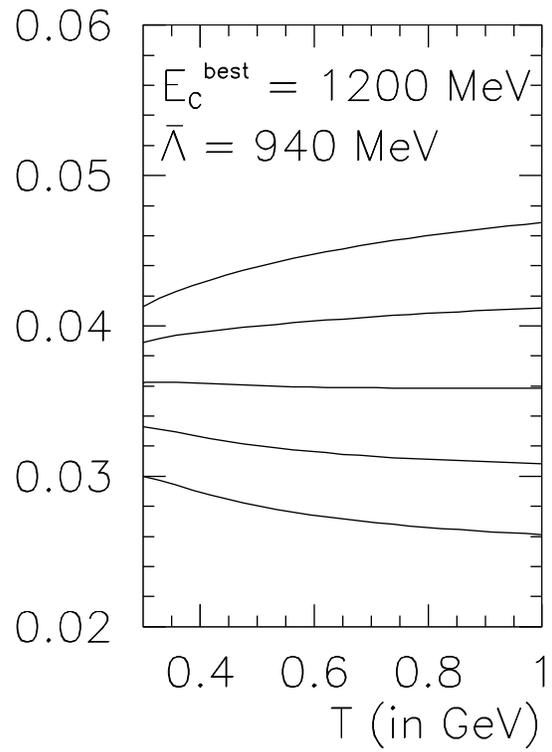
\end{document}